\documentclass[conference,letterpaper]{IEEEtran}
\usepackage{float}
\usepackage{amsmath}
\usepackage{amsfonts}
\usepackage{mathtools}
\usepackage{amssymb}
\usepackage{float}
\usepackage{xcolor}
\usepackage{enumerate}
\usepackage{isomath}
\usepackage{bm}
\usepackage{cite}
\usepackage{graphicx}
\restylefloat{table}
\usepackage{tikz} 
\usepackage{lipsum}
\newcommand\blfootnote[1]{%
  \begingroup
  \renewcommand\thefootnote{}\footnote{#1}%
  \addtocounter{footnote}{-1}%
  \endgroup
}
\usetikzlibrary{arrows, positioning, automata}
\usetikzlibrary{shapes.geometric, arrows}
\def\xfoo#1^#2\relax#3\valign{%
\mathbf{#1}\ifx\valign#2\valign\else^{\mathbf{#2}}\fi}

\def\Sscrm{\pmb{\mathscr{S}}}

\def \Tbf{\mathbf{D}}
\def\Ibf{\mathbf{I}}

\def\Itt{\mathtt{I}}
\def \Sm{\mathrm{S}}
\def\sbf{\mathbf{S}}
\def\Sbf{\boldsymbol{\mathsf{S}}}
\def\xbf{\mathbf{X}}
\def\Xbf{\boldsymbol{\mathsf{X}}}
\def \Xm{\mathsf{X}}
\def \Tm{\mathsf{T}}
\usepackage{amsmath, amssymb, bbm, xspace}
\usepackage{epsfig}
\usepackage{longtable}
\usepackage{color}
\usepackage{mathrsfs}
\usepackage{comment}

\usepackage{courier}
\newtheorem{theorem}{Theorem}[section]

\newtheorem{lemma}{Lemma}[section]

\newtheorem{corollary}[theorem]{Corollary}

\newtheorem{definition}{Definition}[section]

\def\bkE{{\rm I\kern-.17em E}}
\def\bk1{{\rm 1\kern-.17em l}}
\def\bkD{{\rm I\kern-.17em D}}
\def\bkR{{\rm I\kern-.17em R}}
\def\bkP{{\rm I\kern-.17em P}}

\def\bkZ{{\bf{Z}}}

\def\bkE{{\rm I\kern-.17em E}}
\def\bk1{{\rm 1\kern-.17em l}}
\def\bkD{{\rm I\kern-.17em D}}
\def\bkR{{\rm I\kern-.17em R}}
\def\bkP{{\rm I\kern-.17em P}}

\makeatletter
\newcommand{\pushright}[1]{\ifmeasuring@#1\else\omit\hfill$\displaystyle#1$\fi\ignorespaces}
\newcommand{\pushleft}[1]{\ifmeasuring@#1\else\omit$\displaystyle#1$\hfill\fi\ignorespaces}
\makeatother

\def\bkZ{{\bf{Z}}}
\def\b12{(\beta_1,\beta_2)}

\newcounter{example}
\renewcommand{\theexample}{\thesection.\arabic{example}}

\newcounter{remark}
\renewcommand{\theremark}{\thesection.\arabic{remark}}

\def\Xscr{\mathcal{X}}

\def\Ebb{\mathbb{E}}
\newlength{\noteWidth}
\long\def\notes#1{\ifinner
{\tiny #1}
\else
\marginpar{\parbox[t]{\noteWidth}{\raggedright\tiny #1}}
\fi\typeout{#1}}

 \def\notes#1{\typeout{read notes: #1}} %uncomment for final version

\newcommand{\ie}{i.e.\@\xspace} %%% i.e.,
\newcommand{\eg}{e.g.\@\xspace} %%% e.g.,
\def\Ebb{\mathbb{E}}

\def\Pbb{{\mathbb{P}}}

\def\Ibb{{\mathbb{I}}}

\def\ast{\buildrel{t}\over\rightarrow}

\def\exp{\mathop{\hbox{\rm exp}}}

\def\spose#1{\hbox to 0pt{#1\hss}}

\def\text #1{\hbox{\quad#1\quad}}

\def\nthinsp{\mskip -2   mu}

\def\superstar{^{\raise 0.5pt\hbox{$\nthinsp *$}}}
\def\SUPERSTAR{^{\raise 0.5pt\hbox{$*$}}}
\def\lamstarT {\lambda^{\raise 0.5pt\hbox{$\nthinsp *$}T}}

\def\Iscr{{\cal I}}

\def\Xscr{{\cal X}}

\def\non{\nonumber}

\let\forallnew\forall
\renewcommand{\forall}{\forallnew\ }
\let\forall\forallnew

		\def\bkE{{\rm I\kern-.17em E}}
		\def\bk1{{\rm 1\kern-.17em l}}
		\def\bkD{{\rm I\kern-.17em D}}
		\def\bkR{{\rm I\kern-.17em R}}
		\def\bkP{{\rm I\kern-.17em P}}
		\def\bkY{{\bf \kern-.17em Y}}
		\def\bkZ{{\bf \kern-.17em Z}}
		\def\bkC{{\bf  \kern-.17em C}}
{\begin{list}{}%
         {\setlength{\leftmargin}{#1}}%
         \item[]%
}
{\end{list}}

		\def\bsp{\begin{split}}
		\def\beq{\begin{eqnarray}}
		\def\bal{\begin{align*}}
		\def\bc{\begin{center}}
		\def\be{\begin{enumerate}}
		\def\bi{\begin{itemize}}
		\def\bs{\begin{small}}
		\def\bS{\begin{slide}}
		\def\ec{\end{center}}
		\def\ee{\end{enumerate}}
		\def\ei{\end{itemize}}
		\def\es{\end{small}}
		\def\eS{\end{slide}}
		\def\eeq{\end{eqnarray}}
		\def\eal{\end{align*}}
		\def\esp{\end{split}}
		\def\qed{ \vrule height7.5pt width7.5pt depth0pt}  %width4.17pt depth0pt} 

	\def\cp2problem#1#2#3#4{\fbox
		 {\begin{tabular*}{0.9\textwidth}
			{@{}l@{\extracolsep{\fill}}l@{\extracolsep{6pt}}l@{\extracolsep{\fill}}c@{}}
				#1 & & $#4 $ 
			\end{tabular*}}}

		\def\bkE{{\rm I\kern-.17em E}}
		\def\bk1{{\rm 1\kern-.17em l}}
		\def\bkD{{\rm I\kern-.17em D}}
		\def\bkR{{\rm I\kern-.17em R}}
		\def\bkP{{\rm I\kern-.17em P}}
		
		\def\bkZ{{\bf{Z}}}

\newcommand {\beeq}[1]{\begin{equation}\label{#1}}
\newcommand {\eeeq}{\end{equation}}
\newcommand {\bea}{\begin{eqnarray}}
\newcommand {\eea}{\end{eqnarray}}

\def\texitem#1{\par\smallskip\noindent\hangindent 25pt
               \hbox to 25pt {\hss #1 ~}\ignorespaces}

\def\bsp{\begin{split}}
		\def\beq{\begin{eqnarray}}
		\def\bal{\begin{align*}}
		\def\bc{\begin{center}}
		\def\be{\begin{enumerate}}
		\def\bi{\begin{itemize}}
		\def\bs{\begin{small}}
		\def\bS{\begin{slide}}
		\def\ec{\end{center}}
		\def\ee{\end{enumerate}}
		\def\ei{\end{itemize}}
		\def\es{\end{small}}
		\def\eS{\end{slide}}
		\def\eeq{\end{eqnarray}}
		\def\eal{\end{align*}}
		\def\esp{\end{split}}
		\def\qed{ \vrule height7.5pt width7.5pt depth0pt}  %width4.17pt depth0pt} 

\usepackage{amsmath, amssymb, xspace}
\usepackage{epsfig}
\usepackage{longtable}
\usepackage{color}
\usepackage{mathrsfs}
\usepackage{subfig}
\newenvironment{proof}[1][]{{\noindent \textit{ Proof}: }}{\hfill \qed \vspace{3pt}\\ }

\ifCLASSINFOpdf
  \else
 
\fi

\author{\IEEEauthorblockN{Sharu Theresa Jose and Osvaldo Simeone}}
\title{Address-Event Variable-Length Compression for Time-Encoded Data }
\begin{document}
\maketitle
\begin{abstract}
Time-encoded signals, such as social network update logs and  spiking traces in neuromorphic processors, are defined by multiple traces carrying information in the timing of events, or spikes. When time-encoded data is processed at a remote site with respect to the location it is produced, the occurrence of events needs to be encoded and transmitted in a timely
fashion. The standard Address-Event Representation (AER) protocol for neuromorphic chips encodes the indices of the ``spiking" traces in the payload of a packet produced at the same time the events are recorded, hence implicitly encoding the events' timing in the timing of the packet.  This paper investigates the potential bandwidth saving that can be obtained by carrying out variable-length compression of packets' payloads. Compression leverages both intra-trace and inter-trace correlations over time  that are typical in applications such as social networks or neuromorphic computing. The approach is based on discrete-time Hawkes processes and entropy coding with conditional codebooks. Results from an experiment based on a real-world retweet dataset are also provided.
\end{abstract}
\blfootnote{The authors are with King's Communications, Learning, and Information Processing (KCLIP) lab at the Department of Engineering of King’s College London, UK (emails: sharu.jose@kcl.ac.uk, osvaldo.simeone@kcl.ac.uk).
The authors have received funding from the European Research Council
(ERC) under the European Union’s Horizon 2020 Research and Innovation
Programme (Grant Agreement No. 725731). The authors thank Prof. Rajendran (KCL) for useful discussions.}
\vspace{-0.6cm}
%\textit{A full version of this paper is accessible at:}
\section{Introduction}
\begin{figure}[h!]
\captionsetup[subfloat]{oneside,margin={2cm,0cm}}
\centering
\includegraphics[scale=0.31,clip=true, trim = 0.2in  1.8in 0.4in 1.3in]{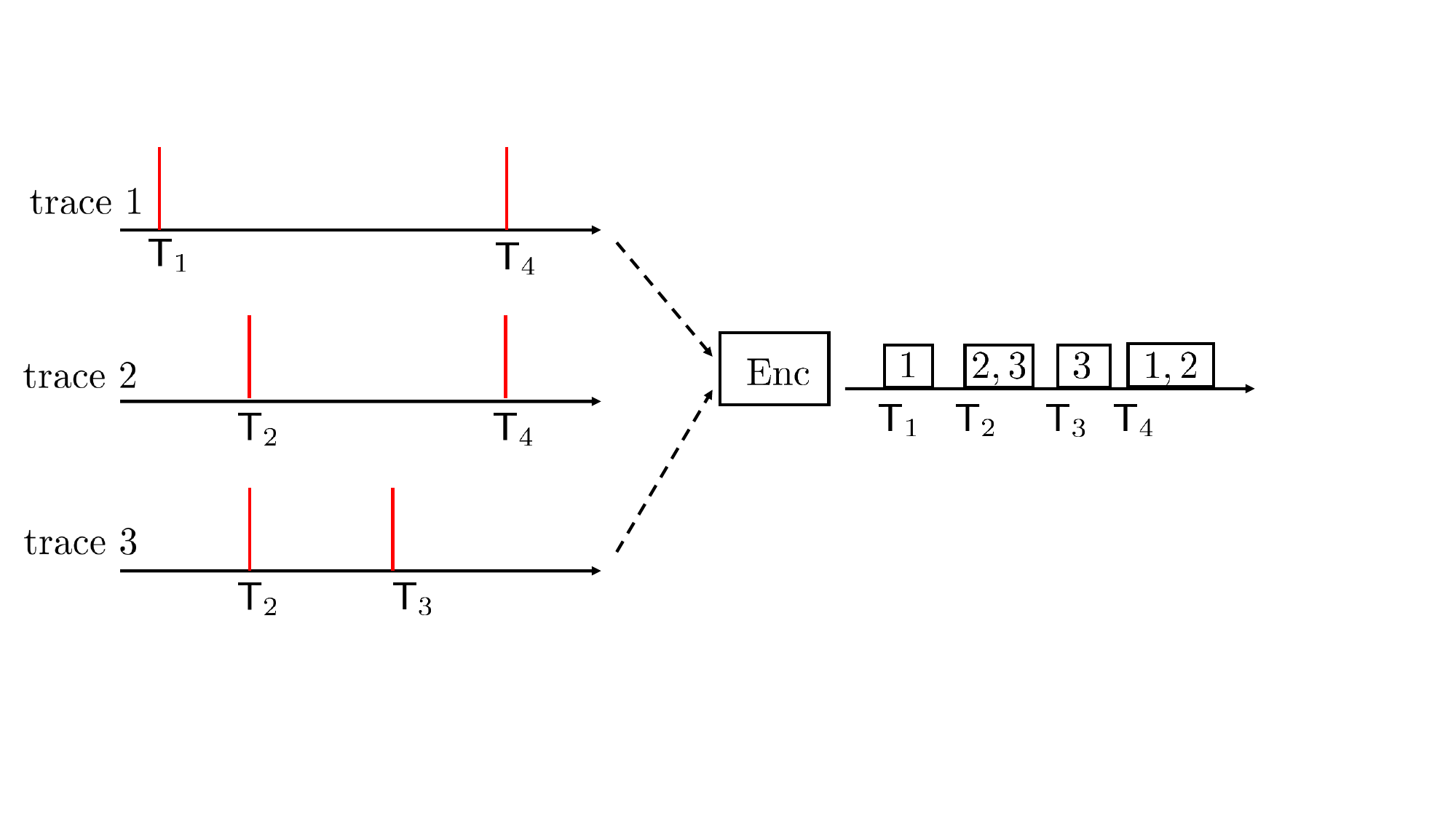} 
\caption{Illustration of the problem of variable-length address-event compression with $N=3$ traces: At each events' occurrence time $\Tm_n$, the encoder outputs a variable-length packet describing the set $\Iscr_n$ of traces producing an event.}\label{fig:multiuserproblem}
\vspace{-0.5cm}
\end{figure}

%$\foo{\mathrm{X}}$ $\mathbf{X}$ $\foo{\textit{X}}$
%$\mathsf{X}$ $\boldsymbol{\mathsf{X}}$ $\boldsymbol{\mathrm{X}}$ $\bm{X}$ $\mathrm{\bm{X}}$ $\bm{\mathrm{X}}$ $\mathcal{\mathsf{I}}$ $\boldsymbol{\textrm{X}}$ $\textrm{X}$
%$\mathboldsans{S}$ $\mathboldsans{s}$  $\mathboldsans{X}$ $\mathboldsans{x}$ $\mathrm{S}$  $\mathbold{X}$  $\Xbf$  $\Sbf$ $S$ $X$
Time-encoded information underlies many data types of increasing importance, such as social network update times \cite{dwyer2007trust}, communication network logs \cite{takahashi2008retrieving}, retweet traces \cite{zhao2015seismic}, wireless activity sensors \cite{endo2012real}, neuromorphic sensors \cite{simeone2019learning,liu2019event}, and synaptic traces from in-brain measurements for brain-computer interfaces \cite{musk2019integrated}. Time-encoded signals are defined by multiple traces, each carrying information in the timing of \emph{events}, also known as spikes (see Figure~\ref{fig:multiuserproblem}). To elaborate on some examples, neuromorphic cameras encode information by producing a spike in response to changes in the sensed environment \cite{liu2019event}; neurons in a Spiking Neural Networks (SNNs) compute and communicate via spiking traces in a way that mimics the operation of biological brains  \cite{simeone2019learning}, \cite{davies2018loihi}; social networks keep logs of update times for all users \cite{dwyer2007trust}; and wireless sensors can measure the activity on given channels as binary (on-off) time-frequency binary maps \cite{endo2012real}. 

When time encoded data is processed at a remote site with respect to the location in which the data is produced, the occurrence of events needs to be encoded and transmitted in a timely fashion. A notable example is given by SNN chips for which neurons are partitioned into several cores, and spikes produced by neurons in a given core need to be conveyed to the recipient neurons in a separate core in order to enable correct processing \cite{davies2018loihi}.
 If a packet encoding the occurrence of one or more events is produced at the same time (within some tolerance) in which the events take place, then timing information is directly carried by the reception of the packet. Therefore, the packet payload only needs to contain information about the identity, also referred to as ``addresses", of  the ``spiking" traces. 
This is the approach taken by the Address Event Representation (AER) protocol, which is the de facto standard for the representation and transmission of time-encoded data in neuromorphic sensors and  implementations of SNNs \cite{mahowald1994analog, boahen2000point, higgins1999multi}.

This paper studies the problem of compressing packets generated by an AER-like protocol for generic time-encoded data. The key idea is that time-encoded traces are typically characterized by strong correlations both over time and across different traces. For instance, the spike timings in biological neural traces reveal excitatory and inhibitory inter-neuron effects \cite{dayan2003theoretical}; and the timing of tweets of different users are correlated within chains of retweets \cite{zhao2015seismic}. These intra- and inter-trace correlations can be harnessed to compress, using variable-length codes, the description of the identity, or addresses, of the event-producing traces at a given time. Our approach is based on discrete-time Hawkes processes \cite{seol2015limit} and entropy coding with conditional codebooks \cite{cover06elements}.

%\subsection{Related Work}
To the best of our knowledge the problem of compression for AER-like protocols has not been studied in the literature. Extensive work has been carried out for the related problems of converting bandlimited or finite-innovation signals into time-encoded information \cite{lazar2004perfect, alexandru2019time}; and of representing point processes as bit streams \cite{lapidoth2015covering}. There are also active lines of research on the definition and learning of statistical models of point processes as sources of time-encoded information \cite{mei2017neural,zhang2019self}; as well as on the estimation of statistical measures of correlations within time-encoded data streams, such as Granger causality or directed information \cite{xu2016learning}. Other related works concern the transmission of time-encoded information over queuing channels that randomly delay input spikes or events \cite{anantharam1996bits,coleman2008rate}. 
%\subsection{Paper Organization}
%
%\textit{Paper Organization}: We formulate the problem in Section~\ref{sec:problemformulation} by introducing a probabilistic model for the correlated sources and by defining the address-event variable-length compression problem. In Section~\ref{sec:distribution}, we discuss key properties of the joint distribution of  the sources that are instrumental in evaluating the achievable variable-length compression rates.  Section~\ref{sec:mainresults} presents the optimal variable-length compression rates. Finally, we conclude in Section~\ref{sec:conclusion} with some discussions on possible future extensions.

\textit{Notation}: 
Throughout this paper, we use upper case  sans-serif letters \eg, $\Xm$, to represent random variables,  
and upper case letters \eg, $X$, to represent realizations of the random variables. We use bold sans-serif letters, \eg, $\Xbf$, to represent random vectors or matrices, and the corresponding upper case letters, \eg, $\xbf$, to denote its realization.  We also use calligraphic upper case letters  \eg, $\Xscr$, to represent random sets, and the upper case typewriter letter, \eg, $\mathtt{X}$, to represent a realization. We use the notation  $[N]=\{1,\hdots,N\}$, and $2^{[N]}$ represents the power set, \ie, the set of all subsets of $[N]$, while $\{\emptyset\}$ represents the empty set. Furthermore, we let $\Tbf$ represent a sparse lower shift matrix with $[\Tbf]_{r,c}=\mathrm{1}(\{r-c=1\})$ where $\mathrm{1}(\cdot)$ is the indicator function which takes value $1$ when `$\cdot$' is true and equals zero otherwise. 
%zeros everywhere except for its lower diagonal elements, and
 We  also use $\mathbf{O}$ to represent an all-zero matrix.

\section{Problem Formulation}\label{sec:problemformulation}
%$\mathrm{S}$ $\mathbf{\mathrm{S}}$ and  $\mathrm{\mathbf{S}}$
In this section, we present the problem formulation by describing first the probabilistic model of the considered sources and then the address-event compression problem.
\subsection{Multivariate Discrete-Time Hawkes Process}
Throughout this paper, as illustrated in Figure~\ref{fig:multiuserproblem}, we consider time-encoded data defined by $N$ discrete-time traces over time index $t=1,2,\hdots$. Each trace records the timings of events from a given source, \eg, tweets from a user or spikes from a neuron. Mathematically, we define
 $\Xm^{(i)}(t) \in \{0,1\}$ as the random variable that takes value $1$ when an event of trace $i$ occurs at time $t$, and is equal to zero otherwise.
%  Let $\{X^{(i)}(t)\}_{t \geq 0}$ represent the event arrival process of trace $i$, where $X^{(i)}(t)=1$ when an event from trace $i$  occurs at time $t$, and equals zero otherwise.
Accordingly, the $N$ dimension row vector
%% consider the scenario where there are multiple sources of events, the arrivals of which are correlated and not independent. Let $n$ represent the number of sources, each producing events belonging to a specific type and $N=[1,
%\hdots n]$ represent the set of all sources. When multiple sources are considered, it is highly likely that at any time instant $t$,  simultaneous occurrence of events of different types occur.  To rightly capture this scenario of multiple correlated sources, we resort to a multivariate Hawkes processes. 
$\Xbf(t)=(\Xm^{(1)}(t), \Xm^{(2)}(t),\hdots, \Xm^{(N)}(t))$ represents the values of all the traces at time $t$.  For each trace $i \in [N]$, we define the time of occurrence of the $k$th event as
\begin{align}
% \Tm^{(i)}_{k}=\min\biggl \lbrace t: \sum_{t'\leq t}\Xm^{(i)}(t')=k\biggr \rbrace,
\Tm^{(i)}_{k}=\min\biggl \lbrace t: \sum_{t'\leq t}\Xm^{(i)}(t')=k\biggr \rbrace,
 \end{align}
and the timing of the $n$th event across all traces as 
\begin{align}
%\Tm_n=\min \biggl \lbrace t: \sum_{i=1}^N\sum_{t' \leq t}\Xm^{(i)}(t')=n \biggr \rbrace. \label{eq:arrivaltime}
\Tm_n=\min \biggl \lbrace t:\sum_{t' \leq t}\mathrm{1}\bigl(\Xbf(t') \neq \mathbf{O}\bigr )=n \biggr \rbrace. \label{eq:arrivaltime}
\end{align} Moreover, we use $\Xbf(1:t-1)=(\Xbf(1),\hdots, \Xbf(t-1))$ to define the (past) history of traces at time $t$.

The arrival of events are generally correlated across time and traces, capturing excitation or  inihibition effects within a trace and among different traces. To account for these effects, each trace $i \in [N]$ is associated with an  intensity function \cite{laub2015hawkes} $\lambda^{(i)}(t|\xbf(1:t-1))$, 
which depends on the history $\xbf(1:t-1)$ of all traces as
%we define  $\Xbf(1:t-1)=(\Xbf(1),\hdots, \Xbf(t-1))$ for the history of traces at time $t$, and associate with each trace $i \in [N]$, an  intensity function $\lambda^{(i)}(t|\Xbf(1:t-1))$ which conditioned on the history, quantifies the excitation or inhibition effects within and across traces. Precisely, the conditional intensity function is defined as
\begin{align}
\lambda^{(i)}(t|\xbf(1:t-1))=\lambda_i+ 
\sum_{j=1}^N \sum_{k: \hspace{0.1cm} T_k^{(j)}<t}\nu_{i,j}(t-T_k^{(j)}) \label{eq:intensity_finitemultiple}.
\end{align}
In \eqref{eq:intensity_finitemultiple}, parameter $ \lambda_i >0$ is a baseline intensity; and each function  $\{\nu_{i,j}(t)\}_{t \geq 1}$,  with $\nu_{i,j}(t)=0$ for $t <1$, is known as the kernel function  for a pair of traces $i,j \in [N]$. As detailed below, a larger intensity $\lambda^{(i)}(t|\xbf(1:t-1))$ implies a larger probability for trace $i$ ``spiking" at time $t$. Therefore, when the $(i,j)$th kernel satisfies $\nu_{i,j}(t')>0$ for some $t'\geq 1$, the occurrence of an event at trace $j$ at time $t-t'$ increases  the intensity of trace $i$ at time $t$. Conversely, a decrease in the intensity of trace $i$ is caused by the same event if the $(i,j)$th kernel satisfies $\nu_{i,j}(t')<0$.
%A decrease in this probability is caused instead when the $(i,j)^{th}$ kernel satisfies $\nu_{i,j}(t')<0$. 
Note also that 
 when $\nu_{i,j}(\cdot)\equiv 0$ for all $i,j \in [N]$, the $N$ traces of events are independent of each other, and each trace $i$ follows a discrete-time Poisson process of rate $\lambda_i$.
A typical example of a kernel is the exponential kernel \cite{laub2015hawkes}, which is defined as
\begin{align}
\nu_{i,j}(t)=\alpha_{i,j}\exp(-t/\beta_{i,j}), \quad \mbox{for}\hspace{0.2cm}  t =1,2,\hdots,  \label{eq:exponential_multi}
\end{align}
%\begin{align}
%\nu(t)=\alpha \exp\biggl(\frac{-t}{\beta}\biggr), \quad t =1,2,\hdots, \tau \label{eq:exponential}
%\end{align}
with parameters $\alpha_{i,j}$ and $\beta_{i,j}>0$ for all pairs of traces $i,j \in [N]$. 
%The positive values of $\alpha_{i,j}$ and $\beta_{i,j}$ imply that the event occurrences of traces is \textit{mutually exciting}, with each occurrence of event of a trace at time $<t$ adds to the intensity of all other traces at time $t$.
%We use the intensity function to define the  probability distribution of the vector $\Xbf(t)$ given the history $\Xbf(1:t-1)$ as detailed next.
% We describe the event arrival process from $N$ correlated sources with finite memory $\tau$ as a $N$-dimensional Hawkes process as defined below.
\begin{definition}[Discrete-Time Multivariate Hawkes Process \cite{laub2015hawkes}]
For a discrete-time $N$-dimensional Hawkes process $\{\Xbf(t)\}_{t \geq 1}$, the probability of vector $\Xbf(t)$ given the history $\Xbf(1:t-1)$ is given as
\begin{align}
&\Pbb[\Xbf(t)=\xbf|\Xbf(1:t-1)=\xbf(1:t-1)]\non \\&=\prod_{i=1}^N \Pbb[\Xm^{(i)}(t)=\xbf^{(i)}|\Xbf(1:t-1)=\xbf(1:t-1)], \label{eq:probability_product}
%\Pbb[\Xbf(t)=\xbf|\Xbf(1:t-1)=\xbf(1:t-1)]=\prod_{i=1}^N \sigma\bigl(\lambda^{(i)}(t|\xbf(1:t-1))\bigr), \label{eq:probability_product}
%\end{align}
%\begin{align}
\\
%&\Pbb[\Xm^{(i)}(t)=1|\Xbf(1:t-1)=\xbf(1:t-1)]\non \\&=\sigma\bigl(\lambda^{(i)}(t|\xbf(1:t-1))\bigr), \label{eq:probability_taumultiple}
&\mbox{where} \quad \Pbb[\Xm^{(i)}(t)=1|\Xbf(1:t-1)=\xbf(1:t-1)]\non\\ &\qquad \qquad=\sigma\bigl(\lambda^{(i)}(t|\xbf(1:t-1))\bigr), \label{eq:probability_taumultiple}
\end{align} with $\sigma(a)=(1+e^{-a})^{-1}$  representing the sigmoid function and $\lambda^{(i)}(t|\xbf(1:t-1))$ being the intensity function in \eqref{eq:intensity_finitemultiple}.
%\begin{align}
%\Pbb[\Xm^{(i)}(t)=1|\Xbf(1:t-1)=\xbf(1:t-1)]&=\sigma\bigl(\lambda^{(i)}(t|\xbf(1:t-1))\bigr), \label{eq:probability_taumultiple}
%\Pbb[X^{(i)}(t)=0|\Xbf(t-\tau:t-1)]&=1-\lambda^{(i)}(t|\Xbf(t-\tau:t-1))
%\end{align}
%for all $i \in [N]$, $\sigma(a)=(1+e^{-a})^{-1}$ is the sigmoid function and $\lambda^{(i)}(t|\Xm(1:t-1))$ is the conditional intensity function for the $i$th trace at time $t$. This is defined as
%\begin{align*}
%\lambda_i+ 
%\sum_{j=1}^N \sum_{k=1}^{\tau}\nu_{i,j}(k) \leq 1 \quad \forall i \in [N].
%\end{align*}
\end{definition}

In this paper, we focus our attention on discrete-time multivariate Hawkes process with finite memory $\tau$. Accordingly, the probability of an event of trace $i$ occurring at any time instant $t$ depends solely on a $\tau-$length history,  \ie, on the \textit{state}
\begin{align}
\Sbf(t)=\Xbf(t-\tau:t-1). \label{eq:state1}
\end{align}
\begin{definition}[Discrete-Time Multivariate $\tau$-Memory Hawkes Process]\label{def:Hawkes_memory}
 A discrete-time  $\tau$-memory $N$-dimensional Hawkes process is defined as in \eqref{eq:probability_product}, \eqref{eq:probability_taumultiple} and \eqref{eq:intensity_finitemultiple}
with kernel $\nu_{i,j}(t)$ satisfying $\nu_{i,j}(t)=0$ for $t>\tau$ and for all $i,j \in [N]$.
%\begin{align}
%\Pbb[X(t)=1|X(t-\tau:t-1)]&=\lambda(t|X(t-\tau:t-1))\label{eq:probability_tau}
%%\Pbb[X(t)=0|X(t-\tau:t-1)]&=1-\lambda(t|X(t-\tau:t-1)),
%\end{align}
%with the conditional intensity function defined as,
% \begin{align}
%\lambda(t|X(t-\tau:t-1))=\lambda_0+ \sum_{t-\tau\leq t_i<t}\nu(t-t_i),\label{eq:intensity_finite}
%\end{align}
%where the baseline intensity $\lambda_0$ and  excitation kernel $\nu(\cdot)$ satisfy $$0\leq \lambda_0+\sum_{k=1}^{\tau}\nu(k)\leq 1.$$
\end{definition}

In the subsequent sections, we use the notation $\Sscrm=\{0,1\}^{\tau N}$ to denote the set of all $2^{\tau N}$ possible states $\Sbf$.
%A typical example of a kernel satisfying Definition~\ref{def:Hawkes_memory} is the truncated exponential kernel \cite{laub2015hawkes} defined as,
%\begin{align}
%\nu_{i,j}(t)=\alpha_{i,j}\exp(-t/\beta_{i,j}), \quad \mbox{for}\hspace{0.2cm}  t =1,2,\hdots,\tau  \label{eq:exponential_multi}
%\end{align}
%\begin{align}
%\nu(t)=\alpha \exp\biggl(\frac{-t}{\beta}\biggr), \quad t =1,2,\hdots, \tau \label{eq:exponential}
%\end{align}
%where $\alpha_{i,j}$ and $\beta_{i,j}>0$ are constants and $\nu_{i,j}(t)=0$ for $t >\tau$.
%When the $(i,j)$ kernel satisfies $\nu_{i,j}(t')>0$ for some $t'\geq 1$, the occurrence of an event at trace $j$ at time $t-t'$ increases the chance of an event at time $t$ for trace $i$. A decrease in this probability is caused instead when the $(i,j)^{th}$ kernel satisfies $\nu_{i,j}(t')<0$. Note also that 
% when $\nu_{i,j}(\cdot)\equiv 0$ for all $i,j \in [N]$, the $N$ traces of events are independent of each other, with each trace of events $i$ following a discrete-time Poisson process of rate $\lambda_i$.
\begin{figure}[h!]
\captionsetup[subfloat]{position=top} 
\begin{center}
\includegraphics[scale=0.32,clip=true, trim = 1.9in  0.4in 2.2in 0.45in]{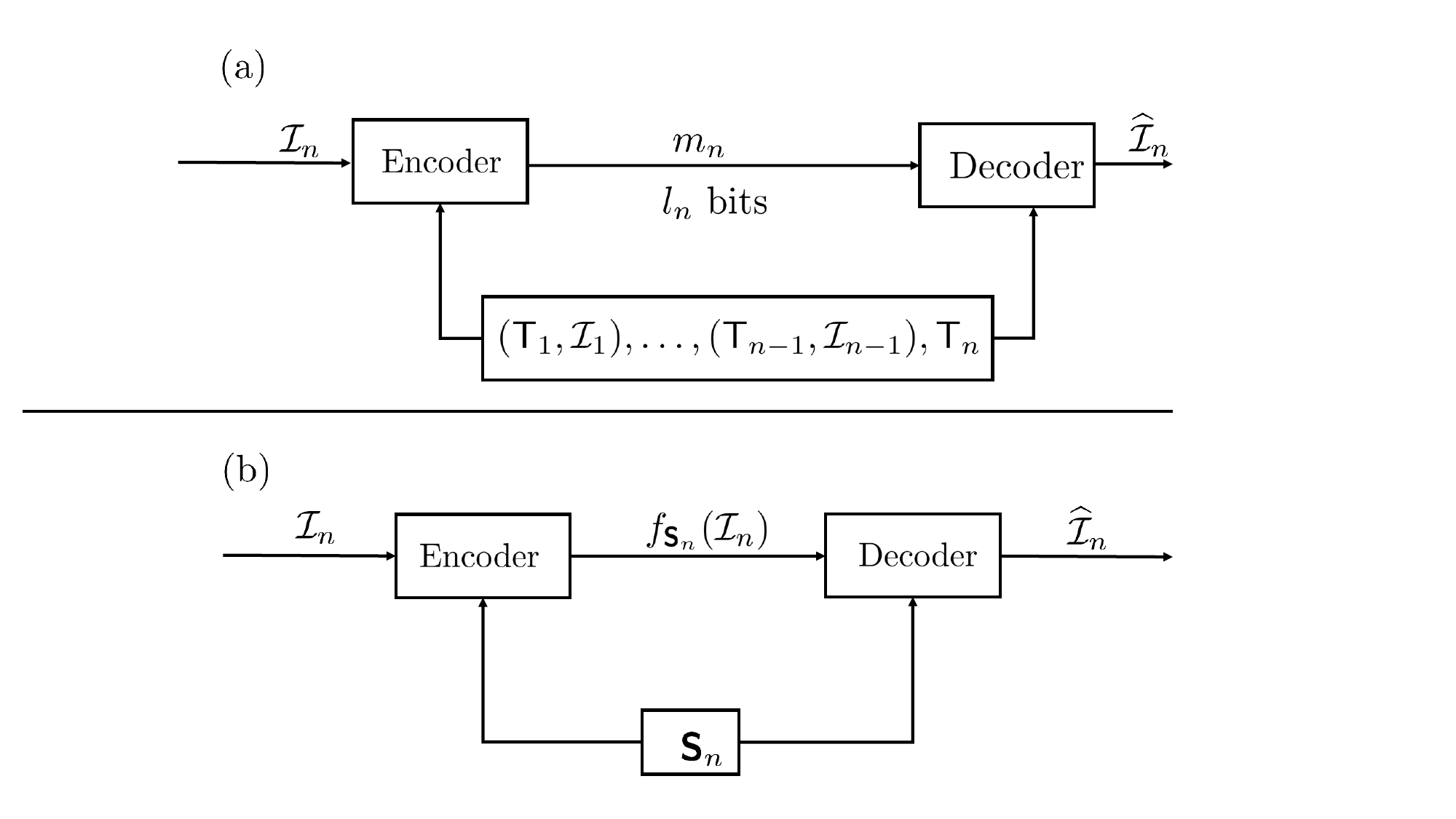} 
\end{center}
%\subfloat[]{\label{fig:encodingfig1}}{\includegraphics[scale=0.33,clip=true, trim = 1.9in  0.4in 0.4in 0.5in]{encodingfig1.pdf} }
%\centering
%\captionsetup[subfloat]{position=top} 
%\subfloat[]{\label{fig:encodingfig2}}{\includegraphics[scale=0.4,clip=true, trim = 1.4in  2.4in 1.3in 2in]{encodingfig2.pdf} }
%\begin{subfigure}{0.75\textwidth}
%\includegraphics[scale=0.4,clip=true, trim = 1.8in  2in 1in 1.8in]{encodingfig1.pdf} 
%\caption{The address-event online variable-length compression problem.}\label{fig:encodingfig1}.
%\end{subfigure}
%\begin{subfigure}{1\textwidth}
%\centering
%\includegraphics[scale=0.4,clip=true, trim = 1.8in  2.4in 1in 1.8in]{encodingfig2.pdf} 
%\caption{The problem setup under consideration.}\label{fig:encodingfig2}.
%\end{subfigure}
\caption{Address-event variable length compression: (a) original problem, and (b) equivalent formulation.}
\vspace{-0.5cm}
\end{figure}

%When $\nu_{i,j}(\cdot)\equiv0$, the Hawkes process defined in Definition~\ref{def:Hawkes_memory} becomes a homogenous discrete-time Poisson process of intensity $\lambda_0$. Finally, the discrete-time Hawkes process in Definition~\ref{def:Hawkes_memory} can also be thought of as a Generalized Linear Model (GLM) \cite{jang2019introduction}.
\subsection{Address-Event Variable-Length Compression}\label{sec:addressevent}
In order to define the compression problem of interest, we first introduce the \textit{address-event process} for the multivariate Hawkes process introduced above.
\begin{definition}[Address-Event Process]
For a discrete-time $\tau-$memory multivariate Hawkes process $\{\Xbf(t)\}_{t \geq 1}$, the address-event process is defined as the random sequence of subsets of indices, or addresses, $\{\Iscr_n\}_{n \geq 1}$, where $\Iscr_n$ is the subset of traces that produce an event at time $\Tm_n$ in \eqref{eq:arrivaltime} as
\begin{align}\Iscr_n=\{i \in [N]: \Xm^{(i)}(\Tm_n)=1\}.\label{eq:indexset}
\end{align}
\end{definition}
We also denote as $\Delta \Tm_{n}= \Tm_{n}-\Tm_{n-1}$ the inter-arrival time between two successive event arrivals from any of the traces, with $\Tm_0=0$.

As illustrated in Figure~\ref{fig:multiuserproblem}, we focus on address-event communication protocols motivated by AER \cite{boahen2000point, sivilotti1991wiring}. Accordingly, at any time $\Tm_n$, the protocol produces a packet describing the index set $\Iscr_n$ of the traces, or addresses, of the events occurring at time $\Tm_n$. Importantly, each packet implicitly carries information about the time $\Tm_n$ in its transmission time, and hence only the subset $\Iscr_n$ needs to be explicitly encoded in the payload of the packet. 
%compression is to represent in a lossless way the index sets $\{\Iscr_n\}_{n\geq 1}$ using packets of short average duration produced at the corresponding times $\{\Tm_n\}_{n \geq 1}$. Importantly, the packets carry implicitly information about the times $\{\Tm_n\}_{n\geq 1}$ in their respective transmission times.

 To elaborate, consider the $n$th arrival time $\Tm_n$. At this point in time, assuming lossless compression, as illustrated in Figure~2(a), the receiver is informed about the previous pairs $\{\Tm_{n'},\Iscr_{n'}\}_{n'\leq n-1}$, and, thanks to the timing of the current packet, also of the current arrival time $\Tm_n$. As a result, the decoder can reconstruct the state vector \begin{align}
\Sbf_n \triangleq \Sbf(\Tm_n)=\Xbf(\Tm_n-\tau:\Tm_n-1).
\end{align} We recall that the state \eqref{eq:state1} summarizes the entire history of the multivariate $\tau$-memory Hawkes Process, yielding the conditional distribution of the set $\Iscr_n$ through \eqref{eq:probability_product} and \eqref{eq:probability_taumultiple}. 
% defines the probability distribution of the set $\Iscr_n$ through \eqref{eq:probability_taumultiple}--\eqref{eq:intensity_finitemultiple}. 
Given that the state $\Sbf_n$ is known to both encoder and the decoder, encoding and decoding functions used to encode and decode the packet produced at any time $\Tm_n$ can be defined to depend on the state $\Sbf_n$ as illustrated in Figure~2(b) and detailed next.
\vspace{-0.2cm}
\begin{definition}
An address-event variable-length code $\{f_{\sbf},g_{\sbf}\}_{\sbf \in \Sscrm}$ consists of
\begin{itemize}
\item an encoding function $f_{\sbf}$ that, for each state $\sbf \in \Sscrm$, maps an index set $\Itt \in 2^{[N]}/\{\emptyset\}$ into a binary string $f_{\sbf}(\Itt)$ of length $l_{\sbf}(\Itt)$ bits;
\item and a decoding function $g_{\sbf}$ that, given the state $\sbf$ and the encoded message, recovers an estimate $\widehat{\Itt}=g_{\sbf}(f_{\sbf}(\Itt)) \in 2^{[N]}/\{\emptyset\}$.
\end{itemize}
\end{definition}
\begin{figure}[h!]
\begin{center}
\includegraphics[scale=0.4,clip=true, trim = 3in  2.6in 3.8in 2.15in]{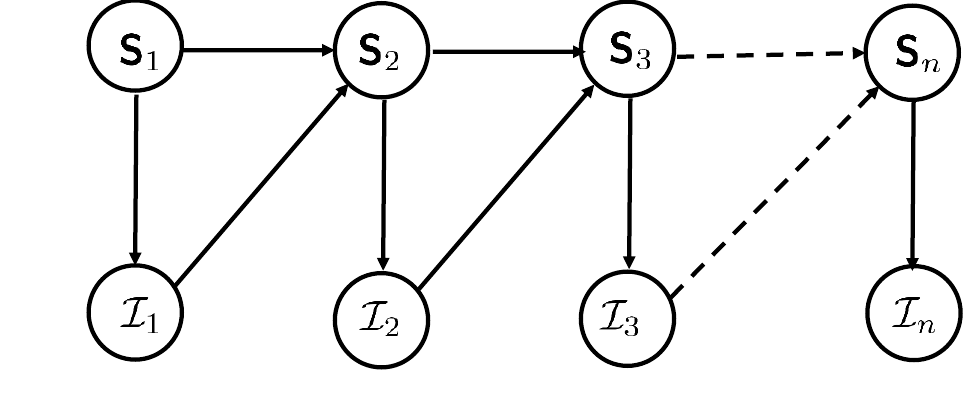} 
\caption{A Bayesian network representation of the address-event process in Theorem~\ref{thm:jointdensity}. Note that, conditioned on the present state $\Sbf_n$,  the index set $\Iscr_n$ is independent of all past index sets $\Iscr_1,\hdots, \Iscr_{n-1}$.}\label{fig:bayesian_addressevent}
\end{center}
\vspace{-0.6cm}
\end{figure}
A code is said to be lossless if the equality $g_{\sbf}(f_{\sbf}(\Itt))=\Itt$ holds for all subsets $\Itt \in 2^{[N]}/\{\emptyset\}$ and states $\sbf \in \Sscrm$. For a lossless code, let $\mathbbm{C}(\sbf)=\{f_{\sbf}(\Itt):\Itt \in 2^{[N]}/\{\emptyset\}\}$ represent the codebook corresponding to state $\sbf$. The expected length of the codewords in $\mathbbm{C}(\sbf)$ is given by the average
\begin{align}
L(\sbf)=\Ebb[l_{\sbf}(\Iscr)|\sbf], \label{eq:averagelength}
\end{align} which is taken with respect to the distribution of $\Iscr$ conditioned on the state $\sbf$. A rate $R$ in \textit{bits per event} is said to be achievable if there exists a lossless code for which the following limit holds
\begin{align}
\limsup_{T \rightarrow \infty} \frac{\sum_{n=1}^{N(T)}L(\Sbf_n)}{N(T)}=R\quad \mbox{a.s},\label{eq:achievableR_multi}
\end{align}
where $N(T)=\max\{n:\Tm_n \leq T\}$ represents the total number of events occurred in the interval $[1,T]$. 
We are interested in characterizing the infimum $R^{*}$ of all achievable rates. 
\section{Distribution of The Address-Event Process}\label{sec:distribution}
In this section, we derive the joint distribution of the address-event process and of the state sequence $\{\Iscr_n,\Sbf_n\}_{n \geq 1}$, as well as the marginal distribution of the states $\{\Sbf_n\}_{n \geq 1}$. These results will be instrumental in characterizing the rates \eqref{eq:achievableR_multi} as detailed in the next section. To start, we define the following conditional probability distributions (CPDs). 

First, we consider the CPD $\Pbb[ \Iscr_n=\Itt|\Sbf_{n}=\sbf]$ of a non-empty subset $\Iscr_n=\Itt \subseteq [N]$ of traces having events' occurrences at time $\Tm_n$ given state $\Sbf_n=\sbf$. This is needed to evaluate the conditional average rate \eqref{eq:averagelength}. By \eqref{eq:probability_product}, the events' occurrences are conditionally independent across traces given the current state, and hence the above CPD evaluates to
%Given the state of the system $\Sm_n=\sbf$, the conditional probability distribution $\Pbb[ \Iscr_n=\Itt|\Sm_{n}=\sbf]$  describes the probability that a non-empty subset $\Itt \subseteq [N]$ of traces have event occurrences at time $T_n$.  The evaluation of the above probability distribution employs the conditional independence of event occurrences of traces at any time given the current state as defined in \eqref{eq:probability_product}. Furthermore, conditioning on the event that the set of indices $\Itt$ is non-empty, the above CPD evaluates to
%
%Key to evaluating this probability distribution is that  recall from \eqref{eq:probability_product} that the probability of event occurrence of a trace  at any time is independent of all other traces given the current state. Together with the conditioning on the event that the set of indices $I$ is non-empty, the above CPD evaluates to
%As a result, the above conditional probability can be evaluated as a product of probabilities that all indices in $I$ have event occurrence at the next time instant and none of other indices in $[N]\I$ have event occurrences. 
%evaluates as
%Given the state of the system $\Sm_n=\sbf$, the probability that a non-empty subset $I \subseteq [N]$ of traces have event occurrences at time $t_n$ evaluates as
\begin{align}
\Pbb[ \Iscr_n=\Itt|\Sbf_{n}=\sbf]=\frac{\prod_{i \in \Itt}\sigma(\lambda_n^{(i)}(\sbf)) \prod_{j \in \Itt^c} (1-\sigma(\lambda_n^{(j)}(\sbf)))}{1-\prod_{j \in [N]}\bigl(1-\sigma ( \lambda_n^{(j)}(\sbf))\bigr)},\label{eq:probability_IS}
\end{align}
where $\Itt^c=[N]/\Itt$ is the complement set of $\Itt$ and the intensity 
\begin{align}
\lambda_n^{(i)}(\sbf)\triangleq \lambda^{(i)}(T_n|\sbf), \quad  \mbox{for} \hspace{0.1cm} i \in [N],
\end{align} is as defined in \eqref{eq:intensity_finitemultiple}. In \eqref{eq:probability_IS}, the numerator corresponds to the probability that only addresses in $\Itt$ have event occurrences at $T_n$, while the denominator evaluates the probability that at least one event occurs at time $T_n$. 

Second, for a given inter-arrival time $\Delta \Tm_{n+1}=\Tm_{n+1}-\Tm_n=\Delta T$, the state $\Sbf_{n+1}$ can be evaluated as a deterministic function of the previous state $\Sbf_n$  and index set $\Iscr_n$ as
\begin{align}
\Sbf_{n+1}&=\eta(\Delta T, \Sbf_n, \Iscr_n) \label{eq:eta}, \\
\mbox{with}\quad \eta(\Delta T, \sbf_n, \Itt)&=\Tbf^{\Delta T -1}\biggl( \Tbf \sbf_n+ \Ibf (\Itt)\biggr), \label{eq:stateevolution}
\end{align}and $\Ibf(\Itt)$ representing a $\tau \times N$ matrix  with $[\Ibf(\Itt)]_{r,c}=\mathrm{1}(\{r=1\} \hspace{0.1cm}\mbox{ and} \hspace{0.1cm} \{c \in \Itt\})$.
%with value ``1" at entries $(1,i)$ with $i \in \Itt$, and  zero entries everywhere else.
 To see why \eqref{eq:eta}-\eqref{eq:stateevolution} hold, note that each time instant with no events causes the state matrix to shift down by one unit with a new all-zero row (no event) added on top, yielding the matrix $\Tbf \Sbf_n$; while events recorded at traces in subset $\Iscr_n=\Itt$ modify the state $\Sbf_{n+1}$ by adding matrix $\Ibf(\Itt)$. 
%  Therefore, if $\Delta \Tm_{n+1}=\Delta T$ given 
% $\Sbf_n=\sbf$ and $\Iscr_n=\Itt$,  none of the traces have events occurring at times $T_n+1,\hdots, T_{n}+\Delta T-1$,  while events in $\Iscr_n$ occur at time $T_n$. The corresponding state at these time instants can be obtained through the downward shifts  of $\Tbf \sbf+\Ibf(\Itt)$ as $\Tbf^{\Delta T-1}(\Tbf \sbf+\Ibf(\Itt))=\eta(\Delta T,\sbf,\Itt)$ for $t=1,\hdots,\Delta T-1$ respectively.
 Using this observation together with the conditional independence of events of traces at a time given the current state as per \eqref{eq:probability_product}, we have the CPD for the inter-arrival times
%$\Pbb[\Delta \Tm_{n+1}=t|\Sm_{n}=\sbf, \Iscr_n=\Itt]$ evaluates to
% which conditioned on $\Sm_n=\sbf$ and $\Iscr_n=\Itt$, the probability that the inter-event arrival time $\Delta \Tm_{n+1}$ equals some value $t$ for $ t\geq 1$ is obtained as
\begin{align}
&\Pbb[\Delta \Tm_{n+1}=\Delta T|\Sbf_{n}=\sbf, \Iscr_n=\Itt]\non \\&=\biggl(\prod_{i \in [N]}\prod_{j=1}^{\Delta T-1}\Bigl(1-\sigma(\lambda_n^{(i)}(\eta(j,\sbf,\Itt)))\Bigr)\biggr) \times \non\\ & \qquad\biggl(1-\prod_{i \in [N]}\bigl(1-\sigma(\lambda_n^{(i)}(\eta(\Delta T,\sbf,\Itt))\bigr) \biggr).\label{eq:probability_deltaS}
%&\Pbb[\Delta \Tm_{n+1}=\Delta T|\Sbf_{n}=\sbf, \Iscr_n=\Itt]\non \\&=\biggl(\prod_{i \in [N]}\prod_{j=1}^{\Delta T-1}\bigl(1-\sigma(\lambda_n^{(i)}(\eta(j,\sbf,\Itt))\bigr)\biggr) \biggl(1-\prod_{i \in [N]}\bigl(1-\sigma(\lambda_n^{(i)}(\eta(\Delta T,\sbf,\Itt))\bigr) \biggr).\label{eq:probability_deltaS}
\end{align}
The first of the product terms in \eqref{eq:probability_deltaS} evaluates the probability that no events from any of the traces occur during the times $
T_n+1 ,
\hdots, 
T_{n}+\Delta T-1$, and the second term similarly computes the probability that at least one event of  a trace occurs at time $T_{n+1}$. From these two observations, 
the joint probability distribution of $\{\Iscr_n, \Sbf_n\}_{n \geq 1}$ is characterized as follows.
\begin{theorem}\label{thm:jointdensity}
The joint distribution of the address-event process and state sequence $\{\Iscr_n, \Sbf_n\}_{n \geq 1}$ factorizes as
\begin{align}
\Pbb[\{\Iscr_n, \Sbf_n\}_{n \geq 1} ]&=\Pbb[\Sbf_1]\prod_{n\geq 1} \Pbb[\Sbf_{n+1},\Iscr_n|\Sbf_n ]\non
\\&=\Pbb[\Sbf_1]\prod_{n\geq 1} \Pbb[\Iscr_n|\Sbf_n ]\Pbb[\Sbf_{n+1}|\Iscr_n,\Sbf_n ], \label{eq:jointdistribution}
\end{align}
where $\Pbb[\Sbf_1]=\Pbb[\Delta \Tm_1|\Sbf_0=\mathbf{O},\Iscr_0=\{\emptyset\}]$ with $\Pbb[\Sbf_1=\mathbf{O}]=1$, the CPD $\Pbb[\Iscr_n|\Sbf_n ]$ is as given in \eqref{eq:probability_IS} and we have the CPD
\begin{align}
&\Pbb[\Sbf_{n+1}=\sbf_{n+1}|\Iscr_n=\Itt,\Sbf_n=\sbf ]\non =\\&\begin{cases}
\Pbb[\Delta \Tm_{n+1}=\Delta T|\Iscr_n=\Itt,\Sbf_n=\sbf ] &\hspace{-1.7cm} \mbox{if}\hspace{0.1cm} \sbf_{n+1}=\eta(\Delta T,\sbf,\Itt)\label{eq:probability_SIS} \\ & \hspace{-1.7cm}\mbox{and} \hspace{0.1cm} 1 \leq \Delta T \leq \tau,
\\1 \hspace{-0.1cm}-\hspace{-0.1cm}\sum_{\Delta T=1}^{\tau}\Pbb[\Delta \Tm_{n+1}=\Delta T|\Iscr_n=\Itt,\Sm_n=\sbf ] \hspace{-0.2cm} & \mbox{if}\hspace{0.1cm} \sbf_{n+1}=\mathbf{O},
\\0 & \mbox{otherwise},
\end{cases} 
\end{align} with  $\Pbb[\Delta \Tm_{n+1}|\Iscr_n=\Itt,\Sbf_n=\sbf ] $ given in \eqref{eq:probability_deltaS}.
\end{theorem}

%\begin{proof}
%We skip the proof here.
%\end{proof}
According to Theorem~\ref{thm:jointdensity}, the joint distribution of the pairs $\{\Iscr_n,\Sbf_n\}_{n \geq 1}$  factorizes according to the Bayesian network \cite{boutilier1996context}  in Figure~\ref{fig:bayesian_addressevent}. As a corollary, it follows that the system states $\Sbf_{n}$ evolves as a Markov chain, as detailed next.
\begin{corollary}\label{cor:Markov_state}
The evolution of system states $\{\Sbf_n\}_{n \geq 1}$ forms a Markov chain such that
$$\Pbb[\Sbf_{n+1}|\Sbf_1,\hdots, \Sbf_{n}]=\Pbb[\Sbf_{n+1}|\Sbf_n],$$
with the transition probability 
\begin{align}
&\Pbb[\Sbf_{n+1}=\sbf_{n+1}|\Sbf_n=\sbf]=\non\\&\begin{cases}
\Pbb[\Delta \Tm_{n+1}=\Delta T,\Iscr_n=\Itt,|\Sbf_n=\sbf ] & \mbox{if}\hspace{0.1cm} \sbf_{n+1}=\eta(\Delta T,\sbf,\Itt)\label{eq:probability_transition} \\ & \mbox{for}\hspace{0.1cm}\Itt \in 2^{[N]}/\{\emptyset\}, \\&\mbox{and}\hspace{0.1cm} 1 \leq \Delta T \leq \tau,
\\
%\sum_{\Itt \in 2^{[N]}/\{\emptyset\}} 
1-\sum_{\Delta T =1}^{\tau}\sum_{\Itt \in 2^{[N]}/\{\emptyset\}} \Pbb[\Delta \Tm_{n+1}\hspace{-0.4cm}&=\Delta T, \Iscr_n=\Itt|\Sbf_n=\sbf ] \\& \mbox{if}\hspace{0.1cm} \sbf_{n+1}=\mathbf{O},
\\0 & \mbox{otherwise}, 
\end{cases}
%\begin{cases} \Pbb[\Delta \Tm_{n+1}=\Delta T,\Iscr_n=\Itt|\sbf] &\mbox{if} \hspace{0.1cm} \sbf_{n+1}=\eta(\Delta T,\sbf,\Itt)\non  \\& \mbox{for} \hspace{0.1cm} \Itt \in 2^{[N]}/\{\emptyset\},  1\leq \Delta T \leq \tau \non \\
%\sum_{\Itt \in 2^{[N]}/\{\emptyset\}} \Pbb[\Delta \Tm_{n+1}> \tau,\Iscr_n=\Itt|\sbf] &\mbox{if}\hspace{0.1cm} \sbf_{n+1}=\mathbf{O}
%\\
%0 &\mbox{otherwise}, 
%\end{cases}
\end{align} 
%\begin{align}
%\Pbb[\Sbf_{n+1}|\Sbf_n]&=\sum_{\Iscr_n \in 2^{[N]}/\{\emptyset\}} \Pbb[\Sbf_{n+1},\Iscr_n|\Sbf_n ] 
%%=\sum_{\Iscr \in 2^{[N]}/\emptyset} \Pbb[\Iscr_n|\Sbf_n ]\Pbb[\Sbf_{n+1}|\Iscr_n,\Sbf_n ]
%\end{align}
where $ \Pbb[\Iscr_n|\Sbf_n ]$ and $\Pbb[\Delta \Tm_{n+1}|\Iscr_n,\Sbf_n]$ are as given in  \eqref{eq:probability_IS} and \eqref{eq:probability_deltaS}.  Moreover, the above Markov chain is irreducible, and there exists a unique $|\Sscrm| 
\times 1$ dimension stationary distribution 
$\boldsymbol{\pi} $ as the solution of the linear system
$
\mathbf{P}\boldsymbol{\pi}=\boldsymbol{\pi}$,
where $\mathbf{P}=\{\Pbb[\Sbf_{n+1}=\sbf'|\Sbf_n=\sbf]\}_{\sbf,\sbf'\in \Sscrm}$ is the transition matrix.
%where $\Pbb[\Iscr_n|\Sbf_n ]$ is as given in \eqref{eq:probability_IS} and  $\Pbb[\Sbf_{n+1}|\Iscr_n,\Sbf_n ]$ is as given in \eqref{eq:probability_SIS}.
\end{corollary}
\begin{proof}
See Appendix~\ref{app:1}.
\end{proof}
\vspace{-0.5cm}
\section{Minimum Rate}\label{sec:mainresults}
In this section, based on the key results reported above, we derive the minimum compression rate $R^{*}$  in bits per event for the address-event compression problem defined in Section~\ref{sec:addressevent}. To this end, we first obtain a general expression for the achievable rate $R$ defined in \eqref{eq:achievableR_multi} as a function of the CPD \eqref{eq:probability_IS} and of the stationary state distribution derived in Corollary~\ref{cor:Markov_state}. 
\begin{lemma}\label{lem:LLN}
The rate $R$ achievable by a
 lossless address-event variable length code $\{f_{\sbf},g_{\sbf}\}_{\sbf \in \Sscrm}$ defined in \eqref{eq:achievableR_multi} equals
%The limiting value of the performance measure evaluates to
\begin{align}
R=\Ebb_{\boldsymbol{\pi}}[L(\Sbf)], \label{eq:limitingvalue}
\end{align} 
where $\boldsymbol{\pi}$ is the stationary distribution of the Markov process $\{\Sbf_n\}_{n\geq 1}$ as defined in Corollary~\ref{cor:Markov_state}.
% with transition probability  given as 
%\begin{align}
%&\Pbb[\delta_{n+1}=j, \Iscr_n=\Iscr|\Sbf_{n}=\sbf] \non \\
%&=\Pbb[ \Iscr_n=\Iscr|\Sbf_{n}=\sbf] \Pbb[\delta_{n+1}=k|\Sbf_{n}=\sbf, \Iscr_n=\Iscr] \non,
%%=\prod_{i \in \Iscr} \sigma(\lambda^{(i)}(\sbf)) \prod_{m \in [N]} \prod_{k=1}^{j-1} \biggl( 1-\sigma \bigl(\lambda^{(m)}(\eta(\sbf,\Iscr))\bigr)\biggr) , \label{eq:probability}
%%&=\exp\biggl ( -\sum_{j=1}^k\lambda(j|s)\Delta\biggr)\lambda(k|s)\Delta+o(\Delta),\label{eq:probability2}
%\end{align} where $I \in 2^{[N]}/\{\emptyset\}$ and 
%
%where $\Iscr^c=[N]/\Iscr$,  $\lambda^{i}(\sbf)= \lambda^{i}(t_n|\sbf)$ is as defined in \eqref{eq:intensity_finitemultiple} and
%\begin{align}
%&\Pbb[\delta_{n+1}=k|\Sbf_{n}=\sbf, \Iscr_n=\Iscr]=\Pbb[\delta_{n+1}=k|\Sh_{n}=T\sbf+I(\Iscr)] \non \\
%&=\biggl(\prod_{i \in [N]}\prod_{j=1}^{k-1}(1-\sigma(\lambda^{(i)}(T^{j-1}\sh_n)))\biggr) \times \non \\&\biggl(1-\prod_{i \in [N]}(1-\sigma(\lambda^{(i)}(T^{k-1}\sh_n))) \biggr) \label{eq:probability_deltaS}
%\end{align}
%where $\sh_n=T\sbf+I(\Iscr)$.
%$s=[x_{t_{n}-1},x_{t_{n}-2},\hdots,x_{t_{n}-\tau}]$,
%\begin{align}
%\lambda^{(i)}(\sbf)=\lambda_i+\sum_{m=1}^N \sum_{k=1}^{\tau}\sbf^{(m)}(k)\nu_{i,m}(k), \label{eq:intensity1}
%\end{align} with $\sbf^{(m)}$ representing the $m^{th}$ column of the matrix.
\end{lemma} 
\begin{proof}
See Appendix~\ref{app:2}.
\end{proof}
%\subsection{Unconstrained Codes}
%We now remove the constraint of prefix-free codes. We note that this constraint is not needed to ensure instantaneous decodability given the separate reception of the packets \cite{alon1994lower}, \cite{szpankowski2011minimum}. In fact, in order to guarantee lossless compression, it is sufficient that all codewords in each codebook $\mathbbm{C}(\sbf)$, $\sbf \in \Sscrm$, are distinct.  Removing the prefix-free constraint, generally improves the compression rate. 
\vspace{-0.05cm}
Based on Lemma~\ref{lem:LLN}, the infimum $R^{*}$ of all achievable rates can be obtained by applying for each state $\sbf \in \Sscrm$ the variable-length code $\mathbbm{C}(\sbf)$ that yields the minimum expected length, as derived in \cite{blundo1996new}.
%To prove the upper bound on $R^{*}$, we adopt a family of unconstrained codes $\{f_{\sbf},g_{\sbf}\}_{\sbf \in \Sscrm}$ designed as in \cite{szpankowski2011minimum}. 
Specifically, for each state value $\sbf \in \Sscrm$, we order the address sets $\Itt \in 2^{[N]}/\{\emptyset\}$ in decreasing order of their conditional probabilities $\Pbb[\Iscr=\Itt|\Sbf=\sbf]$, breaking ties in a lexicographical ordering on $2^{[N]}/\{\emptyset\}$. Codewords are then assigned from the shortest (`0' and `1') to progressively longer ones in this order.
 Mathematically, let \begin{align}
r_{\sbf}:2^{[N]}/\{\emptyset\} \rightarrow \{1,2,\hdots, 2^N-1\} \label{eq:orderingfunction}
\end{align} denote an indexing function such that $r_{\sbf}(\Itt_1)<r_{\sbf}(\Itt_2)$ if $\Pbb[\Iscr=\Itt_1|\Sbf=\sbf]>\Pbb[\Iscr=\Itt_2|\Sbf=\sbf]$ or if $\Pbb[\Iscr=\Itt_1|\Sbf=\sbf]=\Pbb[\Iscr=\Itt_2|\Sbf=\sbf]$ and $\Itt_1$ preceds $\Itt_2$ in lexicographic order. The code $\mathbbm{C}(\sbf)$ assigns to each $\Itt \in 2^{[N]}/\{\emptyset\}$, a codeword of length $l_{\sbf}(\Itt)=\lceil \log_2 ( r_{\sbf}(\Itt)/2 +1)\rceil$ \cite{blundo1996new}. 
\begin{theorem}\label{thm:one2one_multi}
The optimal compression rate $R^{*}$ is given as
%The following bounds on the optimal rate $R^{*}$ hold
\begin{align} R^{*}=\Ebb_{\bm{\pi}}[ \Ebb_{\Iscr}[\lceil \log_2 ( r_{\Sbf}(\Iscr)/2 +1)\rceil]|\Sbf], \label{eq:bounds_one2one}
\end{align}
where function $r_{\Sbf}(\cdot)$ is defined in \eqref{eq:orderingfunction},
% $\Psi : \Real_{+} \rightarrow \Real_{+}$ is the monotonically increasing function
%$$ \Psi(x)=x+(1+x)\log_2(1+x)-x\log_2(x),$$
and the expectation is taken with respect to CPD $   \Pbb[\Iscr|\Sbf] $ defined in \eqref{eq:probability_IS}.
%and $\foo{\pi}$ is the stationary distribution of the Markov chain $\{\Sm_n\}_{n\geq 1}$ as given in Corollary~\ref{cor:Markov_state}.
%
%$\foo{\pi}(s)$ is the stationary distribution of the evolution of system state $S(t)$ with transition probability distribution given by
%\begin{align}
%P_{s,s'}=\begin{cases} \Pbb[\delta=k|s'] &\mbox{if} s=T^ks'+\mathbf{1}_{\tau},  1\leq k\leq \tau-1\\
% \Pbb[\delta\geq \tau|s'] &\mbox{if} s=\mathbf{1}_{\tau}, \\
%0 &\mbox{otherwise}.
%\end{cases}
%\end{align}
\end{theorem}
\begin{figure}[h!]
\begin{center}
\includegraphics[scale=0.4,clip=true, trim = 2.9in 0.9in 2.3in 0.5in]{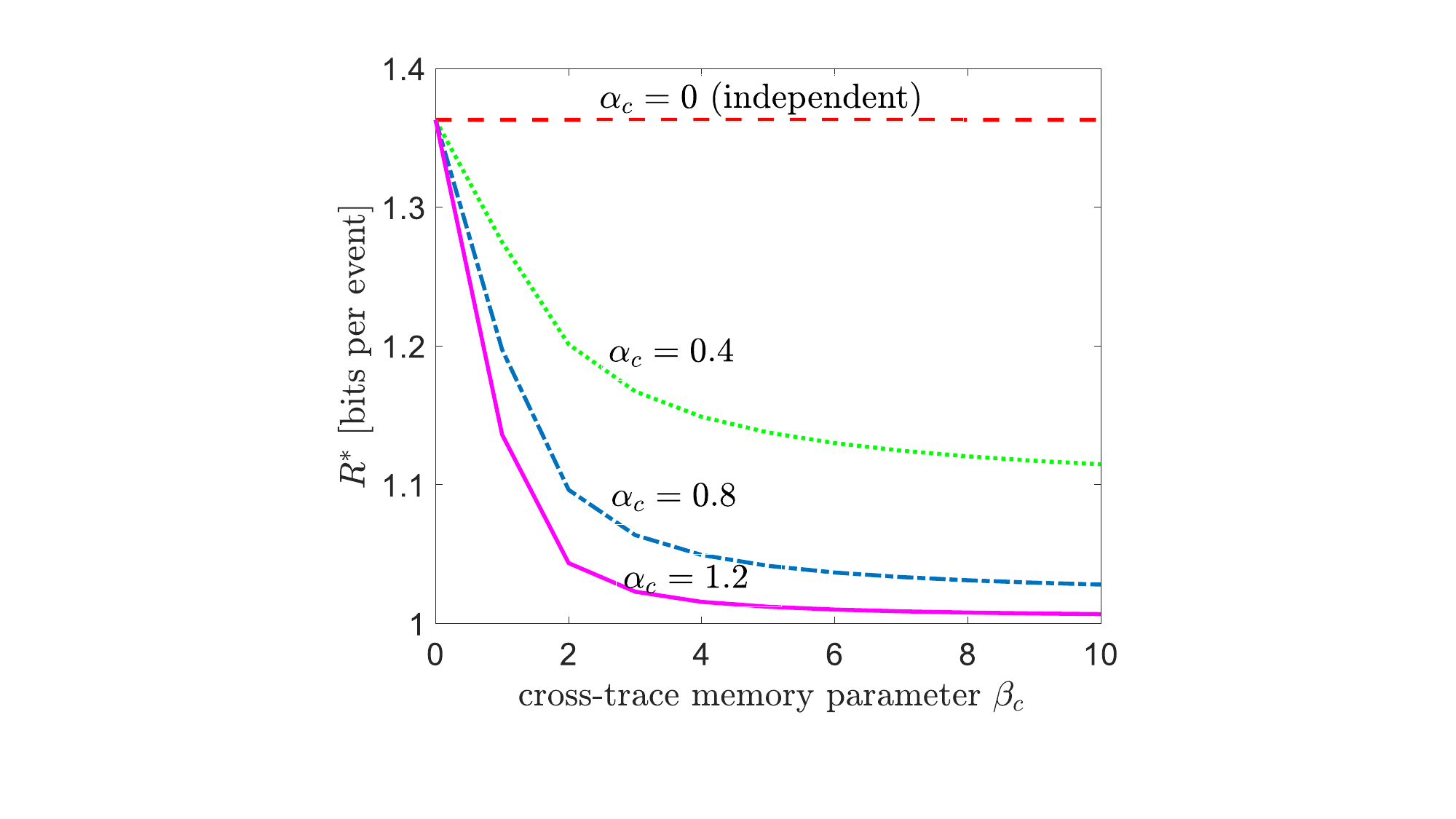} 
\caption{Optimal rate $R^{*}$, in bits per event, versus the cross-trace memory parameter $\beta_c (=\beta_{i,j},  i \neq j$) for $N=3$ traces, $\tau=2$, and the exponential kernel in \eqref{eq:exponential_multi}. We consider varying values of cross-trace correlation parameter $\alpha_c (=\alpha_{i,j}, i \neq j)$, and other parameters are set as $\lambda_1=\lambda_2=\lambda_3=0.4$, $\alpha_{i,i}=0.95$, $\beta_{i,i}=3$ for $i \in [N]$.}\label{fig:intereventarrivalprocess_multi}
\end{center}
\vspace{-0.6cm}
\end{figure}
 Figure~\ref{fig:intereventarrivalprocess_multi} illustrates the optimal rate $R^{*}$ in \eqref{eq:bounds_one2one}, measured in  number of bits per event, for $N=3$ traces  as a function of the cross-trace memory parameter $\beta_c=\beta_{i,j},$ for $ i \neq j$ and $i,j \in [N]$, for the truncated exponential kernel in \eqref{eq:exponential_multi} with $\tau=2$.
 We also vary parameters $\alpha_c=\alpha_{i,j}, i \neq j$,  which determine the strength of the correlation across the two traces of events. Note that when $\alpha_c=0$, the traces are mutually independent, with each following a discrete-time univariate Hawkes process. Other parameters are set as $\lambda_1=\lambda_2=\lambda_3=0.4$, $\alpha_{i,i}=0.95$, $\beta_{i,i}=3$ for $i \in [N]$. A conventional system that does not carry out compression would require a number of bits per event equal to $ \log_2(2^N-1) \approx 2.8$. As seen in Figure~\ref{fig:intereventarrivalprocess_multi}, the proposed variable-length compression scheme  can significantly reduce the average packet length. The rate reduction is particularly pronounced as either $\alpha_c$ increases, so that the correlation between the traces of events strengthens; or as $\beta_c$ increases, which enhances the dependence of present events of a trace on the past occurrences of the other traces.
\section{Experiment on Real-World Dataset}
In this section, we implement the proposed variable-length compression scheme on the retweet dataset  \cite{mei2017neural}. The dataset consists of retweet sequences, each corresponding to the retweets of an original tweet.  Each retweet event in a sequence is marked with the type of user group (`small', `medium' or, `large')
 and with the time (quantized to an integer) elapsed since the original tweet. Accordingly, each sequence can be formatted into $N=3$ discrete-time traces. For our experiments, we sampled 2100 sequences from the data set with 2000  sequences used for training and 100 for testing. The training set is used to fit the parameters $\{\lambda_{i}, \alpha_{i,j}\}_{i,j \in [N]}$ of the Hawkes process with truncated exponential kernel in \eqref{eq:exponential_multi} having fixed memory parameters $\beta_{i,j}=\beta_{i,i}=\beta=40,$ $i,j \in [N]$  and varying $\tau$. This is done by maximizing the training set log-likelihood, where the likelihood of each sequence is given by \eqref{eq:jointdistribution}, via gradient descent log-likelihood estimation \cite{jang2019introduction}. Apart from the described scheme that considers both inter-and intra-trace correlations, for reference, we also consider two simplified strategies:  ``compression with an i.i.d model" which assumes the traces to be independent  (\ie,  $\alpha_{i,j}=0$ for $i,j \in [N]$, $i \neq j$) and memoryless (\ie, $\beta=0$); and ``compression with intra-trace correlation", which assumes independent traces  (\ie,  $\alpha_{i,j}=0$ for $i,j \in [N]$, $i \neq j$) that are allowed to correlate across time. After training, the test sequences are used to evaluate the average number of bits per event, using the trained variable-length code described in Section~\ref{sec:mainresults}.

%the number of followers of the retweeting user and the time elapsed since the original tweet. Grouping the users based on the number of followers, we consider $N=3$ user groups - ``small" (with followers fewer than 120), `medium" (with followers fewer than 1363) and ``large" (comprising of the rest of the users). 
%In line with the mutually-exciting nature of retweets, we considered an exponential kernel for the intensity function, fixing the memory parameters $\tau=50$ and $\beta_{i,j}=\beta_{i,i}=50$. We used maximum log-likelihood estimation on the training sequences to estimate the parameters $\lambda_i, \alpha_{i,j},$  $i,j \in [N]$. 
%%These are then employed to compute the 
%the CPD $\Pbb[\Iscr|\Sbf]$.
% by maximizing the log-likelihood function. Towards this, we considered 100 training sequences and 100 test sequences. The  parameters estimated are then employed to compute the CPD $\Pbb[\Iscr|\Sbf]$. 

\begin{figure}[h!]
\begin{center}
\includegraphics[scale=0.38,clip=true, trim = 2.1in  0.17in 2.8in 0.4in]{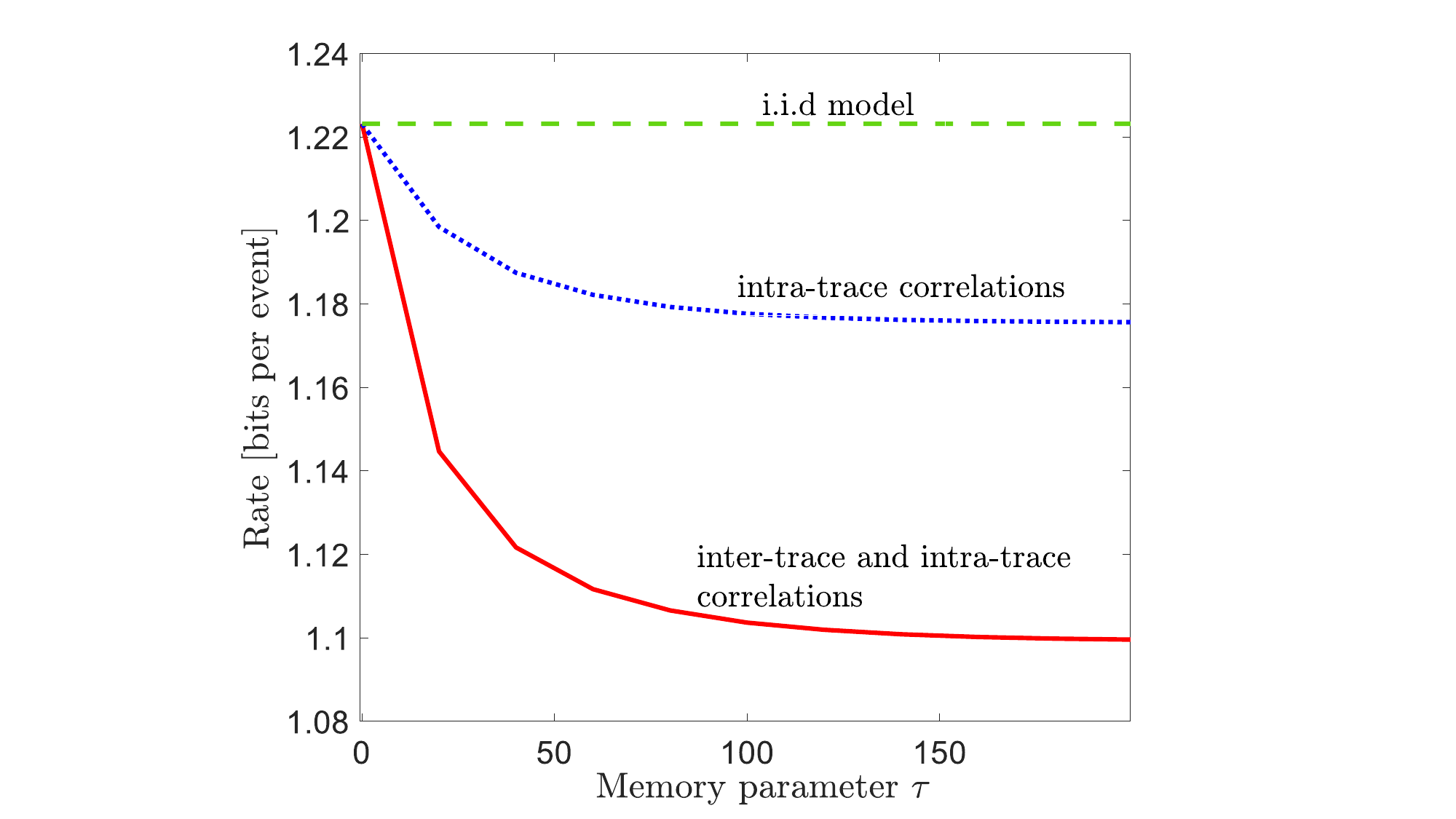} 
\caption{Rate (in bits per event) versus the system memory parameter $\tau$  for the retweet dataset using variable-length compressors trained using Hawkes processes that accounts for no, either, or both intra-trace and inter-trace correlations.}
\label{fig:rate_tau_scenarios}
\end{center}
\vspace{-0.6cm}
\end{figure}

As discussed before, without compression, the required rate would be $\approx 2.8 $ bits per event.  As can be seen from Figure~\ref{fig:rate_tau_scenarios}, compression with an i.i.d model  requires only 1.22 bits per event, a gain of  $57\%$  over no-compression. Further reductions in rates result from compression schemes that assume intra-trace correlation across time, particularly if accounting also for inter-trace correlations.
%wof incorporating intra-trace modelling : Leveraging intra-trace characteristics assuming an i.i.d model with independent  (\ie  $\alpha_c=0$) and memoryless (\ie, $\beta=0$) traces,  Figure~\ref{fig:rate_tau_scenarios} shows a  gain of atleast $57\%$ in compression rate. The compression rate can be improved still by incorporating intra-trace correlation across time for the traces with memory. 
%
% leveraging even a simple intra-trace modelling by considering an i.i.d model with independent traces (\ie  $\alpha_c=0$) which are memoryless compression rate can be improved by considering an i.i.d model with  independent traces ( with $\alpha_c=0$) which are memoryless (\ie, $\beta=0$) ( Leveraging an i.i.d model for the Hawkes process with independent traces For analysis, we consider three scenarioWe implement the variable-length compression scheme  on the testset sequences, and compute the compression rate (averaged over all testset sequences), in bits per event, versus the memory parameter $\tau$, under three different scenarios : `compression with i.i.d model', where the traces are independent (\ie $\alpha_{i,j}=0,$ for $i \neq j$) and memoryless (\ie, $\beta_{i,i}=0$, for $i \in [N]$), `compression with intra-trace correlation', where the traces are independent with intra-trace correlation across time, and `compression with both inter-trace and intra-trace correlation' where the traces are correlated across both traces and time. The results are as shown in Figure~\ref{fig:rate_tau_scenarios}.
\section{Conclusions}\label{sec:conclusion}
This paper has considered the problem of timely communication of generic time-encoded data comprising of multiple traces of timings of events. Adhering to the AER communication protocol, whereby the timing of events is implicitly encoded in the packets' transmission times,  we have proposed the use of variable-length compression to reduce the average packet length by leveraging the correlation of the time-encoded data across time and traces. 
%, we show that considerable gain in compression rates can be achieved over the conventional setting with no compression.
Future works include exploring avenues to reduce the storage complexity of the compression scheme, which requires storing $2^{\tau N}$ codebooks. One promising solution is to employ a universal approximator, such as a neural network, to output the conditional probabilities $\Pbb[\Iscr|\Sbf]$ as a function of the state $\Sbf$. Another interesting extension would be to account for timeout errors resulting in packet dropouts or erasures, which has implications for neuromorphic chips \cite{sakai2019dropout}. 

% While downsampling the states via $\tau$ is an immediate possibility, we hope to further this research direction by considering a novel approach employing a neural network machine to compute the conditional probabilities $\Pbb[\Iscr|\Sm]$, which can then be fed to the encoder/decoder to retrieve the required probabilities. Another extension is to consider the scenario where timeout errors in transmission results in packet dropouts or erasures, which has implications in neuromorphic hardwares \cite{sakai2019dropout}. 
\appendices
\section{Proof of Corollary~\ref{cor:Markov_state}}\label{app:1}
%\begin{proof}[]
It follows from Theorem~\ref{thm:jointdensity} that the CDP at hand is obtained through the marginalization
\begin{align}
&\Pbb[\Sbf_{n+1}=\sbf_{n+1}|\Sbf_n=\sbf]\non \\&=\hspace{-0.3cm}\sum_{\Itt \in 2^{[N]}/\{\emptyset\}}\hspace{-0.5cm}\Pbb[\Iscr_n=\Itt|\Sbf_n=\sbf]\Pbb[\Sbf_{n+1}=\sbf_{n+1}|\Iscr_n=\Itt,\Sbf_n=\sbf ]. \label{eq:equation}
\end{align}
However, from \eqref{eq:stateevolution}, as long as we have $\Delta \Tm_{n+1}\in \{1,\hdots,\tau\}$, the state $\Sbf_{n+1}=\eta(\Delta \Tm_{n+1}, \sbf, \Iscr_n)$ assumes a distinct value for each $\Iscr_n=\Itt \in 2^{[N]}/\{\emptyset\}$.
% Since no two distinct index sets in $2^{[N]}/\{\emptyset\}$ can yield the same state $\Sbf_{n+1}=\sbf_{n+1}$ when $\Delta \Tm_{n+1}\in \{1,\hdots,\tau\}$ given $\Sbf_n=\sbf$,
Therefore, the summation over index sets in \eqref{eq:equation} reduces to the unique index set that produces the state $\sbf_{n+1}$. This results in the transition probability corresponding to the first case in \eqref{eq:probability_transition}. Moreover, given the previous state $\Sbf_n=\sbf$, $\Sbf_{n+1}$ can reach the all-zero state $\mathbf{O}$  with any index set $\Iscr_n=\Itt \in 2^{[N]}/\{\emptyset\}$ when we have $\Delta \Tm_{n+1} >\tau$. Consequently, the summation in \eqref{eq:equation} yields the second case of \eqref{eq:probability_transition}. 

%However, given $\Sbf_n=\sbf$,  \eqref{eq:stateevolution} gives that $\Sbf_{n+1}$ visits a distinct state for each
%$\Iscr_n=\Itt \in 2^{[N]}/\{\emptyset\}$, and for all inter-arrival times $\Delta \Tm_{n+1}\in \{1,\hdots,\tau\}.$
%satisfi es $1 \leq \Delta \Tm_{n+1} \leq \tau$.
%$\Sbf_{n+1}$ can visit  for each $\Itt \in 2^{[N]}/\{\emptyset\}$ and  $1 \leq \Delta \Tm_{n+1} \leq \tau$, \eqref{eq:stateevolution} yields a unique state for $\Sbf_{n+1}$. 
%Consequently, the summation over index sets in \eqref{eq:equation} reduces to the unique index set which produces the state $\sbf_{n+1}$ for given  $\Delta \Tm_{n+1}$. This results in the transition probability corresponding to the first case in \eqref{eq:probability_transition}. Moreover, given the previous state $\Sbf_n=\sbf$, $\Sbf_{n+1}$ can reach the all-zero state $\mathbf{O}$  with any index set $\Iscr_n=\Itt \in 2^{[N]}/\{\emptyset\}$ as long as $\Delta \Tm_{n+1} >\tau$. Consequently, the summation in \eqref{eq:equation} yields the second case of \eqref{eq:probability_transition}. 
Given the previous state $\Sbf_n=\sbf$, there exists $\tau(2^{N}-1)$  distinct states that  $\Sbf_{n+1}$ can visit, excluding the all-zero state. Therefore, the set of all possible states form one closed communicating class, and the chain is irreducible. Moreover, this set is of finite cardinality, \ie, $|\Sscrm|=2^{\tau N}$. Consequently, the Markov sequence $\{\Sbf_n\}_{n \geq 1}$ has a unique stationary distribution.
\section{Proof of Lemma~\ref{lem:LLN}}\label{app:2}
For any lossless code $\{f_{\sbf},g_{\sbf}\}_{\sbf \in \Sscrm}$, the rate \eqref{eq:achievableR_multi} can be written as 
\begin{align*}
\frac{\sum_{n=1}^{N(T)}L(\Sbf_n)}{N(T)}=\sum_{\sbf \in \Sscrm}L(\sbf)\frac{\sum_{n=1}^{N(T)}\Ibb\{\Sbf_n=\sbf\}}{N(T)},
\end{align*}
where $\sum_{n=1}^{N(T)}\Ibb\{\Sbf_n=\sbf\}/N(T)$ measures the fraction of visits to state $\sbf$ by the Markov process $\{\Sbf_n\}_{n\geq 1}$ in a total of $N(T)$ event occurrence times. It then follows from the strong law of large numbers that the limit \begin{align*}
 &\limsup_{T \rightarrow \infty} \frac{\sum_{n=1}^{N(T)}\Ibb\{\Sbf_{n}=\sbf\}}{N(T)}\rightarrow \pi(\sbf) \quad \mbox{a.s},\end{align*}
holds, which yields \eqref{eq:limitingvalue}. 
\bibliographystyle{IEEEtran}
\bibliography{ref}
\end{document}